%% file: phaseretrieval.tex
\documentclass[conference,onecolumn,11pt]{IEEEtran}  

\def\thesistitle{Stable Recovery from the Magnitude of Symmetrized Fourier Measurements }
\def\thesisauthor{Philipp Walk}
\include{packages}
\include{HeaderShamaiElsevier}

\begin{document}

\title{\thesistitle}
\author{%
\IEEEauthorblockN{Philipp Walk}%
\IEEEauthorblockA{Lehrstuhl f{\"u}r Theoretische Informationstechnik\\
Technische Universit{\"a}t M{\"u}nchen\\
Theresienstrasse 90, 80290 M{\"u}nchen\\
Email: philipp.walk@tum.de}
\and
\IEEEauthorblockN{Peter Jung}
\IEEEauthorblockA{Lehrstuhl f{\"u}r Informationstheorie und \\ Theoretische Informationstechnik\\
Technische Universit{\"a}t Berlin\\
Einsteinufer 25, 10587 Berlin\\
Email: peter.jung@mk.tu-berlin.de}
}

\maketitle

\begin{abstract}
In this note we show that stable recovery of complex-valued signals $\vx\in\C^n$ up to global sign can be achieved
from the magnitudes of $4n-1$ Fourier measurements when a certain \emph{symmetrization and zero-padding} is
performed before measurement ($4n-3$ is possible in certain cases). For real signals, symmetrization itself is linear and
therefore our result is in this case a statement on uniform phase retrieval. Since complex conjugation is involved, such measurement procedure
is not complex--linear but recovery is still possible from magnitudes of linear measurements on, for example,
$(\text{Re}(\vx),\text{Im}(\vx))$.

%
%
\end{abstract}

\section{Introduction}\label{sec:phaseretrieval}

Recovering a signal from intensity (magnitude) measurements is known as the \emph{phase retrieval problem}. This problem
has a long history beginning in the $70$'s by  \namen{Gerchberg} and \namen{Saxton} \cite{GS72} and later by
\namen{Fienup} \cite{Fie78}, who gave explicit reconstruction algorithms for the phase from magnitude Fourier
measurements. Since the magnitude of a linear measurement can not distinguish between numbers of unit modulus, stability
and injectivity for such measurements can only hold up to a global phase resp.  sign, i.e.  up to a factor $e^{i\ome}$
resp. $\pm 1$. One of the challenging tasks in phase retrieval is to determine the necessary and sufficient number of
linear measurements for stability or injectivity.  For example, \namen{Candes} et.al.  \cite{CSV12}
have shown stable reconstruction of any $n-$dimensional complex-valued signal from the magnitude of $\Omi(n\log n)$ linear
Gaussian-random measurements. A more principal result from \namen{Balan} et al. in \cite{BCE06} shows that a generic
frame exists with injectivity at $4n-2$  measurements. Moreover, they could give a fast reconstruction
algorithm in \cite{BBCE07}.  In a recent result \cite{BCMN13}, \namen{Bandeira} et al. conjecture that $4n-4$ linear
measurements are necessary for injectivity.  However, a practical construction and implementation of measurements 
at this limiting number seem to be rather hard, but it serves as a theoretical bound.

More recently,  non-linear or interference--based  approaches are considered to provide 
unique phase reconstruction. For example, \namen{Wang} \cite{Wan13} presented a method where interference with a known signal $\vy\in\C^n$ 
helps to recover a signal $\vx\in\C^n$ up to a global sign from only $3n$ Fourier measurements $|\Fmatrix(\vx+\ome\vy)|^2$
where $\ome\in\C$ is a root of unity.
For \emph{real} $k$--sparse signals, \namen{Eldar} and \namen{Mendelson} \cite{EM12} established  stable recovery
from $\Omi(k\log (en/k))$ subgaussian random measurements with high probability.  
A very recent result \cite{EFS13} from \namen{Ehler, Fornasier} and \namen{Sigl} even extends this to the complex case
and provides an explicit reconstruction algorithm.  \namen{Lu} and \namen{Vetterli} also use sparsity for spectral
factorization of real valued impulse responses  \cite{LV11}. Moreover, they also give a reconstruction algorithm.  
A recent result by \namen{Wang} and \namen{Xu} \cite{WX13} states injectivity for $k-$sparse
complex-valued signals from $4k-2$ generic measurements as long as $k<n$.
Unfortunately, so far (to the authors knowledge) there doesn't exists a constructive or deterministic frame providing a recovery or even stable recovery.

Here, we will show a concrete measurement procedure allowing  stable recovery 
of any vector $\vx\in \C^n$ with $x_0\in \R$ up to global sign from 
magnitudes of $4n-3$ linear measurements. The measurements can implemented as linear mappings on, for example,
$(\text{Re}(\vx),\text{Im}(\vx))$ or $(\vx,\bar{\vx})$.
We want to stress the fact, that our measurements are
\emph{not complex--linear}, since we perform a non-linear symmetrization on the signal
to obtain a symmetric
auto-convolution, allowing magnitude measurements from $4n-3$ linear Fourier measurements. However, this will have
implications on certain (compressive) signal processing tasks since such type of measurements occur prior 
to I/Q--down conversion into a suitable complex baseband model.
To prove stability for magnitude Fourier measurements, we will use our result in \cite{WJ12b} for the $(s,f)-$sparse
zero-padded circular convolution. In view of sparsity, zero padding can also be seen as a particular structured sparse
signal subclass in $\C^{4n-3}$.

\section{Circular Convolutions, Correlations and the RNMP}
Let $(\Fmatrix)_{kl}:=n^{-\frac{1}{2}}\exp(i2\pi\frac{kl}{n})$ be the 
$k,l\in\{0,\dots, n-1\}$ elements of the
$n\times n$ discrete Fourier transform (DFT) matrix. 
If dimension of a matrix is important it also will occur as a subscript, i.e. here $\Fmatrix=\Fmatrix_n$.
As well--known, $\Fmatrix$ is unitary and $\vGam:=\Fmatrix^2$ denotes time--reversal given by its action 
$\vGam\cdot(x_0,\dots,x_{n-1})^T:=(x_0,x_{n-1},\dots,x_1)^T$. In particular $\vGam$ is an involution, i.e.
$\vGam^2=\Fmatrix^4=1$. The circular convolution $\sum_{l=0}^{n-1} x_{l } y_{k\ominus l}$ 
($\ominus$ and $\oplus$ mean $\pm$ modulo $n$) of two vectors $\vx,\vy\in\C^n$ 
is a symmetric bilinear mapping given as:
\begin{align}
   \vx\circledast \vy= \sqrt{n}\Fmatrixa(\Fmatrix\vx \odot \Fmatrix\vy)=\vy\circledast\vx
   \label{eq:ccdef}
\end{align}
and $\vx\circledast \vx$ is called (circular) auto--convolution.
Similarly, the circular correlation $\sum_{l=0}^{n-1} x_{l } \bar{y}_{k\oplus l}$ is defined as
$\vx\ostar \vy:=\vx\circledast\vGam\bar{y}$ and we have that Fourier transform 
of the \emph{auto--correlation}:
\begin{align}
   \Fmatrix(\vx\ostar \vx)=\sqrt{n} \Fmatrix \vx\odot \Fmatrix\vGam\cc{\vx}=
   \sqrt{n}\Fmatrix \vx \odot \cc{\Fmatrix \vx}=
   \sqrt{n}|\Fmatrix\vx|^2
   \label{eq:acc:magnitude}
\end{align}
is given as the squared magnitudes of the Fourier transform of $\vx$.
Furthermore, $(\vS^i)_{kl}=\delta_{k\ominus i,l}$ denotes the elements of $i$th power
of the unit shift operator $\vS$.

In \cite{WJ12b} and \cite{Wal13} we have established a stability statement 
for zero-padded sparse circular convolutions.
Let $\text{supp}(x):=\{i:x_i\neq 0\}$ be the support of a vector in the canonical basis
and $\Sigma^n_k:=\{x\in\C^n\,:\,|\text{supp}(x)|\leq k\}$ be the $k$--sparse vectors.
We have the following result on the \emph{restricted norm multiplicativity property} (RNMP) 
for the circular convolution of sparse zero--padded signals (see \cite{WJ12b} for the general definition):
%
\begin{theorem}[RNMP for circular convolutions, \cite{WJ13a,Wal13}]
   \label{cor:dcryi}
  Let $s,f,n\in \N$ with $s \leq f\leq n$. Then there exist a constant $\alp_{n'}>0$ with 
  $n'=n'(s,f,n):=\min\{\tn(s,f),n\}$, such that  for all  $\vx\in\Sigma_s^n,\vy\in\Sigma^n_f$ 
  it holds
  \begin{align}
    \alp_{n'} \Norm{\vx}\Norm{\vy}\leq \Norm{(\vx,\zero) \circledast (\vy,\zero)}\leq
    \sqrt{s}\Norm{\vx}\Norm{\vy},\label{eq:circrnmp}
  \end{align}
  where $(\vx,\zero),(\vy,\zero)\in\C^{2n-1}$ denotes the vectors padded by $n-1$ zeros.
\end{theorem}
%
Note, that for sufficiently small $s$ and $f$ the constant $\alpha_{n'}$ depends \emph{solely} on the sparsity and not 
on the ambient dimension $n$ \cite{Wal13}\footnote{Our first approach on an explicit formula for $\tn(s,f)$ in
  \cite{WJ13a} has been corrected in \cite{Wal13}}. Furthermore, without additional restrictions, zero padding is necessary to obtain a lower
bound strictly greater than zero (see for example also \cite{WJ13a} for an explicit example here).  In fact, Theorem
\ref{cor:dcryi} is a statement on regular convolutions.  However, it is natural to expect also a bound without zero
padding in prime dimension.  Moreover, from $\vx\circledast\vy=\vS\vx\circledast\vS\vy$ follows that \eqref{eq:circrnmp}
holds whenever the zeros are contained in a cyclic block of size $n-1$.

%
\section{Recovery from the Magnitude of Symmetrized Fourier Measurements}

Our contribution is motivated by the framework given in \cite{WJ12b} on bilinear maps. Let
$B(\vx,\vy)$ be a symmetric bilinear map  and denote
its diagonal part by $A(\vx)=B(\vx,\vx)$. Obviously there holds the binomial--type formula:
\begin{equation}
   A(\vx_1)-A(\vx_2)
   =B(\vx_1,{\vx}_1)-B(\vx_2,{\vx}_2) + B(\vx_1,\vx_2)-B(\vx_1,\vx_2)= B(\vx_1-\vx_2,{\vx_1+\vx_2}) 
   \label{eq:binom}
\end{equation}
establishing that such $\vx_1$ and $\vx_2$ can be (stable) distinguished modulo global sign on the basis of $A(\vx_1)$
and $A(\vx_2)$ whenever $B(\vx_1-\vx_2,\vx_1+\vx_2)$ is well--separated from zero. More precisely, such a condition is
given by the RNMP (given in \eqref{eq:circrnmp} for the special case $B(\vx,\vy)=\vx\circledast \vy$ to be considered
here).  Since $B(\vx,\vy)=\vx\circledast \vy$ is symmetric it follows from \eqref{eq:binom}, Theorem \ref{cor:dcryi} and
the results in \cite{WJ12b} that each (zero padded) $s$--sparse $\vx$ for sufficiently large $n$ can be stable recovered
modulo global sign from $O(s\log n)$ compressive i.i.d. subgaussian (and suitable generalizations based on
concentration properties) samples of its circular auto--convolution (which itself can have a sparsity up to $s^2$).
However, more important is the estimation of $\vx$ based on measurements on its auto--correlation $\vx\ostar \vx$. In
particular, for Fourier measurements this corresponds to the observation of intensity, see \eqref{eq:acc:magnitude}.
But, circular correlation $\vx\ostar \vy$ is only symmetric when $\vx=\vGam\bar{\vx}$ (if and only if and the same also
for $\vy$).  In general, a symmetrization $\Sym\colon \C^n \to \C^{2n-1}$ is therefore necessary here:
\begin{align}
   \Sym( \vx):
   =(\overbrace{x_0,x_1,\dots,x_{n-1}}^{=\vx},\overbrace{\bar{x}_{n-1},\dots,\bar{x}_1}^{=:\vx_-^\circ})^T
   \label{eq:sym}
\end{align}
Let us stress the fact, that the symmetrization map is linear only for \emph{real} vectors $\vx$ since complex
conjugation is involved. On the other hand, $\Sym$ can obviously be written as linear map on vectors like $(\text{Re}(\vx),\text{Im}(\vx))$
or $(\vx,\bar{\vx})$. 
Now, for $x_0=\bar{x}_0$ the symmetry condition $\Sym(\vx)=\vGam\cc{\Sym(\vx)}$ is fullfilled
(note that here $\vGam=\vGam_{2n-1}$):
\begin{align}
   \Sym(\vx)=\begin{pmatrix}{\vx}\\ \cc{\vx^\circ_-} \end{pmatrix}
   =\vGam\begin{pmatrix}\cc{\vx}\\ \vx^\circ_- \end{pmatrix}
   =\vGam\cc{\begin{pmatrix}\vx\\\cc{\vx_-^\circ} \end{pmatrix}}
   =\vGam\cc{\Sym(\vx)}.
   \label{eq:involutioninv}
\end{align}
Let us abbreviate therefore $\C^n_0:=\{\vx\in\C^n\,:\,x_0\in\Reals\}$.  Thus, for $\vx,\vy\in\C^n_0$, circular
correlation of symmetrized vectors is symmetric and agrees with the circular convolution.  To apply  Theorem \ref{cor:dcryi}
we define the \emph{zero-padded symmetrization} (first zero padding, then symmetrization) $\Symz : \C^n \to \C^{4n-3}$
by:
\begin{equation}
   \Symz(\vx):=\Sym \begin{pmatrix}\vx\\
      \zero_{n-1} \end{pmatrix}
   \label{eq:s0},
\end{equation}
\begin{theorem}\label{thm:phaseretrieval}
  Let $n\in\N$, then $\tn=4n-3$ absolute-square Fourier measurements of zero padded symmetrized vectors in $\C^{\tn}$,
  given by \eqref{eq:s0}, are stable up to a global sign for $\vx\in\C^{n}_0$, i.e. for all $\vx_1,\vx_2\in \C^n_0$ it
  holds
  \begin{align}
     \Norm{|\Fmatrix\Symz (\vx_1)|^2- |\Fmatrix\Symz(\vx_2)|^2}
     &\geq  c\Norm{\Symz(\vx_1-\vx_2)}\Norm{\Symz(\vx_1+\vx_2)}
     \label{eq:stablequadratic}
  \end{align}
 with $c=c(\tn)=\alp_{\tn}/\sqrt{\tn}>0$ and $\Fmatrix=\Fmatrix_{\tn}$.
\end{theorem}
%
Note that $2\lVert \vx\rVert^2\geq \lVert \Symz(\vx)\rVert^2=\lVert \vx\rVert^2+\lVert \vx^\circ_-\rVert^2\geq\lVert \vx\rVert^2$.  Thus,
$\Symz(\vx)=0$ if and only if $\vx=0$ and the stability in distinguishing $\vx_1$ and $\vx_2$ up to a global sign
follows from the RHS of \eqref{eq:stablequadratic} and reads explicitly as:
\begin{equation}
   \Norm{|\Fmatrix\Symz (\vx_1)|^2- |\Fmatrix\Symz(\vx_2)|^2}
   \geq  c \Norm{\vx_1-\vx_2}\Norm{\vx_1 + \vx_2}.
\end{equation}
\begin{proof}
   For symmetrized vectors $\Symz(\vx)$,  auto--convolution agrees with auto--correlation and
   we get from \eqref{eq:acc:magnitude}:
   \begin{align}
     \Fmatrix(\Symz(\vx)\circledast\Symz(\vx))=\sqrt{\tn}\Betrag{\Fmatrix\Symz(\vx)}^2.
   \end{align}
   Putting things together we get for every $\vx\in\C^n_0$:
   \begin{align}
     \sqrt{\tn}\Norm{|\Fmatrix\Symz (\vx_1)|^2 \!-\!\Betrag{\Fmatrix\Symz(\vx_2)}^2}
      \!&= \Norm{\Fmatrix(\Symz(\vx_1,\vx_1)-\Symz(\vx_2,\vx_2))}\\
      \text{$\Fmatrix$ \small is unitary}\ra &=\Norm{\Symz(\vx_1,\vx_1)-\Symz(\vx_2,\vx_2)}
      \overset{\eqref{eq:binom}}{=}\Norm{\Symz (\vx_1 -\vx_2) \!\circledast\! \Symz (\vx_1 +\vx_2)}\notag\\
      \text{\small Theorem \ref{cor:dcryi}}\ra 
      &\geq\!\alp_{\tn}\Norm{\Symz (\vx_1 -\vx_2)}\cdot\Norm{\Symz (\vx_1 +  \vx_2)}
   \end{align}
   In the last step we use that Theorem \ref{cor:dcryi} applies whenever the non--zero entries are contained in
   a cyclic block of lenth $2n-1$.
\end{proof}
In the \emph{real case} \eqref{eq:stablequadratic} is equivalent to a \emph{stable linear embedding} in $\R^{4n-3}$ up
to global sign (see here also \cite{EM12} where \namen{Eldar} and \namen{Mendelson} used the $\ell^1-$norm on the left
side) and therefore this is an \emph{explicit phase retrieval statement} for \emph{real} signals. Recently, stable
recovery also in the complex case up to global phase from the same number of subgaussian measurements has been achieved by
\namen{Ehler} et al. in \cite{EFS13}.  Such results hold with exponential high probability whereby our
result is deterministic.  But, since $\Symz$ is \emph{not complex--linear} Theorem
\ref{thm:phaseretrieval} can not directly be compared with the usual complex phase retrieval results.  On the other hand, 
such an approach can now indeed distinguish the complex phase by the Fourier measurements and symmetrization provides injectivity for magnitude
Fourier measurements up to global sign.
To get rid of the odd definition $\C_0^n$ one could symmetrize (and zero padding)  $\vx\in\C^n$ also by:
\begin{align}
   \Symz(\vx):=(\zero_{n},x_0\dots x_{n-1},\bar{x}_{n-1}\dots\bar{x}_0,\zero_{n-1})^T\in\C^{4n-1}\label{eq:s2}
\end{align}
again satisfying $\Symz(\vx)=\vGam_{4n-1}\cc{\Symz(\vx)}$ at the price of two 
further dimensions. Hence, we also have:
\begin{corrolary}
   Let $n\in\N$, then $\tn=4n-1$ absolute-square Fourier measurements of zero padded and symmetrized vectors given by
   \eqref{eq:s2} are
   stable up to a global sign for $\vx\in\C^n$, i.e. for all $\vx_1,\vx_2\in \C^n$ it holds
   \begin{align}
      \Norm{|\Fmatrix \Symz (\vx_1)|^2- |\Fmatrix\Symz(\vx_2)|^2}
      \geq  2 c \Norm{\vx_1-\vx_2}\Norm{\vx_1 + \vx_2} 
      \label{eq:stablequadratic2}
   \end{align}
   with $c=c(\tn)=\alp_{\tn}/\sqrt{\tn} >0$ and $\Fmatrix=\Fmatrix_{\tn}$.
\end{corrolary}

The proof of it is along the same steps as in \thmref{thm:phaseretrieval}.
The direct extension to sparse signals as in \cite{WJ12a} seems to be difficult since randomly chosen Fourier 
samples do not provide a sufficient measure of concentration property without further randomization.

\section{Conclusion}
In this note we have shown stable recovery (up to global sign) of a signal $\vx$ from magnitude measurements on the Fourier transform
of its symmetrization $\Symz(\vx)$. For real signals this  procedure is linear and establishes therefore a phase retrieval method.
However, also in the complex case this has practical relevance and system design implications when considering 
linear measurements on $(\text{Re}(\vx),\text{Im}(\vx))$ (or $(\vx,\bar{\vx})$). Our result is deterministic
and uniform, i.e. it guarantees recovery up to global sign for any vector $\vx\in\C^n$. Finally, the constant in the stability result
depends only on the sparsity of $\vx$ indicating a possible further reduction of the number of observations in the Fourier
domain also in this case.

\bibliographystyle{IEEEtran}
\bibliography{jabref_philipp_utf2}

\end{document}

%% file: packages.tex
\usepackage[automark]{scrpage2} 
\usepackage{amsmath}
\usepackage{mathrsfs}						                   
\usepackage{MnSymbol}
\usepackage{amsxtra}           
\usepackage{amsthm}						                      
\usepackage{bbold}           
\usepackage{accents}
\usepackage{times}           
\usepackage[english]{babel}
\usepackage{textcomp}                                               
\usepackage[dvips]{color}                                        
\usepackage{ifpdf}
\ifpdf
  \usepackage[pdftex]{graphicx}
  \DeclareGraphicsExtensions{.pdf}
\else
  \usepackage{graphicx}
  \usepackage{epsfig}
  \DeclareGraphicsExtensions{.eps}
\fi

\usepackage[utf8]{inputenc}                                         
\usepackage[T1]{fontenc}

\usepackage{booktabs}                                               
\usepackage{longtable}
\usepackage{algorithm}
\usepackage{algorithmic}
\usepackage{textcomp}        
\usepackage{caption}
\usepackage{subcaption}
\usepackage[all]{xy}
\usepackage{xytree}
\usepackage{verbatim}
\usepackage{srcltx}                                                 

\usepackage{multirow}

\usepackage{rotating}   

\usepackage{enumerate}
\usepackage[nice]{nicefrac}

%% file: HeaderShamaiElsevier.tex
\newcommand{\Ao}{{\ensuremath{\mathcal{A}^\circ}}}
\newcommand{\Aoz}{{\ensuremath{\mathcal{A}^\circ_{z}}}}
\newcommand{\Symz}{{\ensuremath{\mathcal{S}_z}}}
\newcommand{\Symo}{{\ensuremath{\mathcal{S}^\circ}}}
\newcommand{\Symoz}{{\ensuremath{\mathcal{S}^\circ_{z}}}}

\newcommand{\alp}{\ensuremath{\alpha}}

\newcommand{\vGam}{\ensuremath{{\boldsymbol\Gamma}}}

\newcommand{\ome}{\ensuremath{\omega}}

\newcommand{\tn}{\ensuremath{\tilde{n}}}

\newcommand{\for}{\ensuremath{\quad{\text{for}\quad}}}

\newcommand{\Omi}{{\ensuremath{\mathcal{O}}}}

\newcommand{\Sym}{{\ensuremath{\mathcal{S}}}}

\newcommand{\ko}{{\ensuremath{\mathbb K}}}
\newcommand{\C}{{\ensuremath{\mathbb C}}}

\newcommand{\R}{{\ensuremath{\mathbb R}}}
\newcommand{\N}{{\ensuremath{\mathbb N}}}

\newcommand{\Fmatrix}{{\ensuremath{\mathbf F}}}
\newcommand{\Fmatrixa}{{\ensuremath{\mathbf F}^*}}

\newcommand{\ra}{\rightarrow}

\newcommand{\vx}{{\ensuremath{\mathbf x}}}

\newcommand{\vy}{\ensuremath{\mathbf y}}

\newcommand{\zero}{{\ensuremath{\mathbf 0}}}

\newcommand{\vS}{\ensuremath{\mathbf S }}                         
\newcommand{\vh}{\ensuremath{\mathbf h }}                         

\newcommand{\thmref}[1]{Satz~\ref{#1}}

\newcommand{\noi}{\noindent}

\ifx\definition\undefined
\newtheorem{definition}{Definition}         
\fi
\ifx \@definition \@empty
\newtheorem{definition}[theorem]{Definition}   
\fi

\newtheorem{corrolary}{Corrolary} 
\ifx\conjecture\undefined
\newtheorem{conjecture}{Conjecture}         
\fi
\ifx\theorem\undefined
\newtheorem{theorem}{Theorem}         
\fi
\ifx\lemma\undefined
\newtheorem{lemma}{Lemma}
\fi
\ifx\question\undefined
\newtheorem{question}{\color{red}{Question}}         
\fi
\ifx\proposition\undefined
\newtheorem{proposition}[theorem]{Proposition} 
\fi
{
\comment\noi}%
{\endcomment}

{\par\noindent{\em Beweis\/}.}%
{\hspace*{\fill}{\qed}\vspace{1ex}\par}
{\par\noindent{\em Proof\/}.}%
{\par}

{\hspace*{\fill}{}\vspace{1ex}\par}
{\par\vspace{1.5ex}\noindent{\em Remark\/}.}
{\par\vspace{1.5ex}}
\ifx\remark\undefined
\newenvironment{remark}%
{\par\vspace{1.5ex}\noindent{\em Remark\/}.}
{\par\vspace{1.5ex}}
{\par\vspace{1.5ex}\noindent{\em Example\/}. }
{\par\vspace{1.5ex}}
{\noi\vspace{0.5ex}\small}
{\vspace{0.5ex}\par\normalsize}

\newcounter{Examplecount}
{\color{gray}}
{\vspace{0.5ex}\par\normalsize}

{\renewcommand{\labelenumi}{(\roman{enumi})}\begin{list}{\labelenumi}
{\usecounter{enumi}\setlength{\labelwidth}{1.5cm}\setlength{\topsep}{0.3cm}\setlength{\itemsep}{-3pt}}}
{\end{list}}
{\renewcommand{\labelenumi}{(\arabic{enumi})}\begin{list}{\labelenumi}
{\usecounter{enumi}\setlength{\labelwidth}{1.5cm}\setlength{\topsep}{0.3cm}\setlength{\itemsep}{-3pt}}}
{\end{list}}
{\renewcommand{\labelenumi}{$\bullet$}\begin{list}{\labelenumi}
{\setlength{\labelwidth}{1.5cm}\setlength{\topsep}{0.3cm}\setlength{\itemsep}{-2pt}}}
{\end{list}}

\makeatletter
\newcommand{\set}[2]{\ensuremath{%
\setbox0=\hbox{\ensuremath{#2}}
\dimen@\ht0
\advance\dimen@ by \dp0
\left\{\left.#1\rule[-\dp0]{0pt}{\dimen@}\;\right|\;#2\right\} }}
\makeatother

\newcommand{\Betrag}[1]{\ensuremath{ \left|#1\right| }}

\newcommand{\Norm}[1]{\ensuremath{ \left\|#1\right\| }}

\newcommand{\cc}[1]{{\ensuremath{\overline{#1}}}} 

\newcommand{\namen}[1]{{\textsc{#1}}}           

\ifx \@paragraph \@empty
\makeatletter
\renewcommand\paragraph{\@startsection
{paragraph}{4}{\z@}{-3.5ex plus-1ex minus-.2ex}%
{1.3ex plus.2ex}{\normalfont\itshape}}

\fi

\DeclareUnicodeCharacter{04B4}{\CYRTETSE}
\newcommand{\Reals}{\mathbb{R}}

\renewcommand{\thmref}[1]{Theorem~\ref{#1}}

\definecolor{gray}{rgb}{0.3,0.3,0.3}
{\color{black}}                      
{\color{black}}


%% file: phaseretrieval.bbl
\begin{thebibliography}{10}
\providecommand{\url}[1]{#1}
\csname url@samestyle\endcsname
\providecommand{\newblock}{\relax}
\providecommand{\bibinfo}[2]{#2}
\providecommand{\BIBentrySTDinterwordspacing}{\spaceskip=0pt\relax}
\providecommand{\BIBentryALTinterwordstretchfactor}{4}
\providecommand{\BIBentryALTinterwordspacing}{\spaceskip=\fontdimen2\font plus
\BIBentryALTinterwordstretchfactor\fontdimen3\font minus
  \fontdimen4\font\relax}
\providecommand{\BIBforeignlanguage}[2]{{%
\expandafter\ifx\csname l@#1\endcsname\relax
\typeout{** WARNING: IEEEtran.bst: No hyphenation pattern has been}%
\typeout{** loaded for the language `#1'. Using the pattern for}%
\typeout{** the default language instead.}%
\else
\language=\csname l@#1\endcsname
\fi
#2}}
\providecommand{\BIBdecl}{\relax}
\BIBdecl

\bibitem{GS72}
R.~W. Gerchberg and W.~O. Saxton, ``A practical algorithm for the determination
  of phase from image and diffraction plane pictures,'' \emph{Optik}, vol.~35,
  p. 237–246, 1972.

\bibitem{Fie78}
J.~R. Fienup, ``Reconstruction of an object from the modulus of its fourier
  transform using a support constraint,'' \emph{JOSA A}, vol.~3, pp. 27--29,
  1978.

\bibitem{CSV12}
E.~J. Candes, T.~Strohmer, and V.~Voroninski, ``Phaselift: Exact and stable
  signal recovery from magnitude measurements via convex programming,''
  \emph{Communications on Pure and Applied Mathematics}, vol.~66, pp.
  1241--1274, 2012.

\bibitem{BCE06}
R.~Balan, P.~Casazza, and D.~Edidin, ``On signal reconstruction without
  phase,'' \emph{Applied and Computational Harmonic Analysis}, vol.~20, pp.
  345--356, 2006.

\bibitem{BBCE07}
R.~Balan, B.~G. Bodmann, P.~Casazza, and D.~Edidin, ``Fast algorithms for
  signal reconstruction without phase,'' in \emph{Proc. SPIE 6701}, 2007.

\bibitem{BCMN13}
A.~S. Bandeira, J.~Cahill, D.~G. Mixon, and A.~A. Nelson, ``Saving phase:
  Injectivity and stability for phase retrieval,'' \emph{EURASIP Journal on
  Applied Signal Processing}, 2013.

\bibitem{Wan13}
Y.~Wang, ``Minimal frames for phase retrieval,'' in \emph{Workshop of phaseless
  recovery}, 2013.

\bibitem{EM12}
Y.~Eldar and S.~Mendelson, ``Phase retrieval: Stability and recovery
  guarantees,'' \emph{arxiv}, pp. 1--39, 2012.

\bibitem{EFS13}
M.~Ehler, M.~Fornasier, and J.~Siegl, ``Quasi-linear compressed sensing,''
  \emph{SIAM: Multiscale Modeling and Simulation}, vol. submitted, 2013.

\bibitem{LV11}
Y.~Lu and M.~Vetterli, ``Sparse spectral factorization: Unicity and
  reconstruction algorithms,'' 2011.

\bibitem{WX13}
Y.~Wang and Z.~Xu, ``Phase retrieval for sparse signals,'' \emph{arxiv}, 2013.

\bibitem{WJ12b}
P.~Walk and P.~Jung, ``Compressed sensing on the image of bilinear maps,'' in
  \emph{{ISIT}}, 2012, pp. 1291 -- 1295.

\bibitem{Wal13}
P.~Walk, ``Analysis of convolutions with finite support,'' Ph.D. dissertation,
  TU München, 2013.

\bibitem{WJ13a}
P.~Walk and P.~Jung, ``On a reverse $\ell_2$--inequality for sparse circular
  convolutions,'' in \emph{{ICASSP}}, 2013.

\bibitem{WJ12a}
------, ``Approximation of {L}{\"o}wdin {O}rthogonalization to a {S}pectrally
  {E}fficient {O}rthogonal {O}verlapping {PPM} {D}esign for {UWB I}mpulse
  {R}adio,'' \emph{EURASIP Journal on Applied Signal Processing}, vol.~92, pp.
  649--666, 2012.

\end{thebibliography}
